\documentclass[12pt]{iopart}

\usepackage{iopams}
\usepackage{graphicx}                             % DVI/PS output
\usepackage{color}
\usepackage{multirow}

\begin{document}

\title[]{Energy Calculations of The Realistic Quantum Slab}

\author{G. Bilge\c{c} Aky\"uz$^1$, K. Akg\"ung\"or$^1$, S. \c{S}akiro{\u g}lu$^1$,\\ A. Siddiki$^{2,3}$ and {\.I}. S{\"o}kmen$^1$}
\address{$^1$Dokuz Eyl\"ul University, Physics Department, T\i naztepe Campus, 35100 \.Izmir, Turkey}
\address{$^2$Istanbul University, Physics Department, Beyazit Campus, \.Istanbul, Turkey}
\address{$^3$Department of Physics, Harvard University, Cambridge, MA 02138, USA }
\ead{afifsiddiki@gmail.com}
\begin{abstract}
We calculated the total energy of a semiconductor quantum dot which
is defined by the trench gate method. In our calculation we used a
recently developed energy functional called ``orbital-free energy
functional". We compared the total energies obtained by Thomas-Fermi
approximation, orbital-free energy functional and standard
local-density approximation for the square quantum slab geometry. We
have seen that this newly developed energy functional is numerically
very efficient, superior to the Thomas-Fermi approximation and is in
good agreement with the local-density approximation for two
different sizes of quantum dot systems.
\end{abstract}

\pacs{73.21.La , 71.10.-w, 31.15.B }
%\noindent{\it Keywords}: Quantum Dots, Theories of many electron systems, Thomas-Fermi approximation
\vspace{2pc}

\submitto{\JPCM}
\maketitle

\section{Introduction\label{intro}}
Applications of the two-dimensional (2D) quantum dots (QDs) and the
rectangular quantum slabs (QSs) constitutes basic components of
nanoelectronics and nanotechnology. Especially, the 2D electronic
systems have attracted concern since the beginning of the
developments in nanoelectronics and nanotechnology
\cite{Jacak98:Book}. One of the most important problems in physics
is handling of the electron-electron interactions concerning
many-body systems. The Density Functional Theory (DFT)
\cite{Parr_Gross-book-springer} offers a solution to the problem
with a good accuracy with a reasonable computational cost.
Furthermore, the DFT with two dimensional-local density
approximation has become a standard method for the electronic
structure calculations of the semiconductor QDs
\cite{Reimann02:1283}.

The most critical point of DFT is to describe the exchange and the
correlation functionals in many-body
systems\cite{Sakiroglu10:012505}. Therefore new investigations in
describing exchange and correlations of 2D density by functionals
are important and lead to remarkable results
\cite{Rasanen10:195103,Rasanen10:115108,Gori-Giorgi09:166402,Constantin08:016406}.
The local-density approximation (LDA) provides reasonably good
results for the many-body systems, where the total density is the
sole input variable instead of electronic orbitals as in the DFT.
However, the number of electrons treated numerically is limited,
since the Kohn-Sham (KS) scheme in DFT requires the computation of
single particle KS orbital for the kinetic energy calculation
\cite{Pittalis09:165112}. An alternative theory, called orbital-free
DFT \cite{Wang00:117,Ligneres05:springer,Watson00:128} is more
convenient than the traditional DFT. In contrast, this approach is
more complicated in constructing an accurate energy functional
especially for the three dimensional (3D) systems. An important
example of orbital-free DFT is the traditional Thomas-Fermi
approximation (TFA) that is analyzed in 2D \cite{Lieb95:10646} and
is applied successfully in calculations of electronic structures
\cite{Siddiki03:125315}. However, the TFA treats the
electron-electron (e-e) interaction only classically. Thus it is not
a good approximation for small electron densities, \emph{i.e.} in
the strong interaction regime.

In this study, we use an orbital-free energy functional (OFEF)
\cite{Pittalis09:165112} to calculate the total energy of
interacting electrons in two dimensions for different number of
particle systems considering the QDs. In Ref.
\cite{Pittalis09:165112}, the authors report quite consistent
results with the local-density approximation and provide
considerable improvements over the TFA. They also report that
utilizing the orbital-free functional for 2DES is numerically very
efficient by the virtue of orbital free calculation scheme and due
to the calculation of Hartree integral.

In our study we calculate the energies of a quantum dot defined
in quantum slab geometry, by using OFEF, TFA and LDA. These quantum
dots are obtained by trench-gating \cite{Siddiki07:045325}.
\section{Model}
\begin{figure}[h!] \centering
\includegraphics[width=0.6\columnwidth]{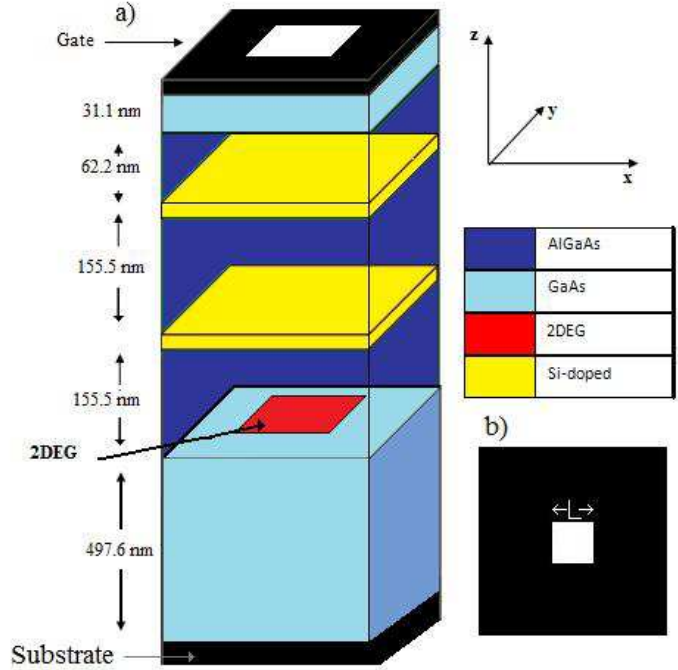}
\caption{(a) The silicon doped heterostructure, the 2DES is formed
at the interface of the GaAs/AlGaAs denoted by red region. The
metallic gates are deposited on the surface. (b) Top view of the
gate model of quantum slab, the edge width (L) of the square dot
Sample-I and Sample-II are $1.73~\mu m$ and $1.11~\mu m$,
respectively.} \label{fig:slab}
\end{figure}

The realistic modeling of two-dimensional systems (2DES) relies on
solving the 3D Poisson equation for given boundary conditions, set
by the GaAs/AlGaAs heterostructure and surface patterns, as shown in
Fig.\ref{fig:slab}-(a). The heterostructure consists of metallic
surface gates determining the charge and the potential
distributions, the 2DES, and $\delta$-doped Silicon layer which
provides electrons to the 2DES. The electron gas is formed at the
interface of the GaAs/AlGaAs hetero-junction. The number of
electrons, $N$ and the average electron density, $n_{\rm
el}$ are determined by the donor density $n_0$ and the metallic
gates. To obtain the potential and the charge distribution of the
system, first, the gate voltage $V_g$ and the donor density are
fixed and next the Poisson equation is solved self-consistently. For
the solution, we used the code which based on a fourth-order
algorithm operating on a square grid. This code is suitable for
different boundary conditions which is applied and tested in
previous studies \cite{Aslan-Tez,Aslan08:125423}.
As an illustrating example in Fig.~\ref{fig:slab}, the semiconductor
surface is partially covered by a patterned gate or trench gate.
Negatively charged metallic gate is biased with -1.0 V depicted by
the black area as shown in Fig.~\ref{fig:slab}-b. We used two different
samples, namely Sample-I and Sample-II, with different slab's area
corresponding to different density parameters $\rm r_s$. The
dimensions of the square slabs are $1.73 \rm \mu m$ and $1.11 \rm
\mu m$ considering Sample-I and Sample-II, respectively. The bare
confinement potential of the system can also be obtained
analytically \cite{Siddiki07:045325, Davies94:4800}, starting from
the lithographically defined pattern.
The un-patterned surface is taken to be pinned to the mid-gap of the
heterostructure and is set to be the reference potential,
$V_{unpat}(\mathbf{r},0)=0$. The external potential $V_{ext}(x,y,z)$
is calculated for $z=0$ plane by the solution of Laplace equation
$\nabla^2V_{ext}(\mathbf{r})=0$ and next we seek for a solution to
the Poisson equation
$\nabla^2V_{ext}(\mathbf{r})=4\pi\rho(\mathbf{r})$, where
$\rho(\mathbf{r})$ is the total charge density, considering the
boundary condition $\partial V_{ext}/\partial z->0$ as $z->\infty$.
\section{Numerical Procedure and Results}
\label{}
\begin{figure}[h!] \centering
\includegraphics[width=0.7\columnwidth]{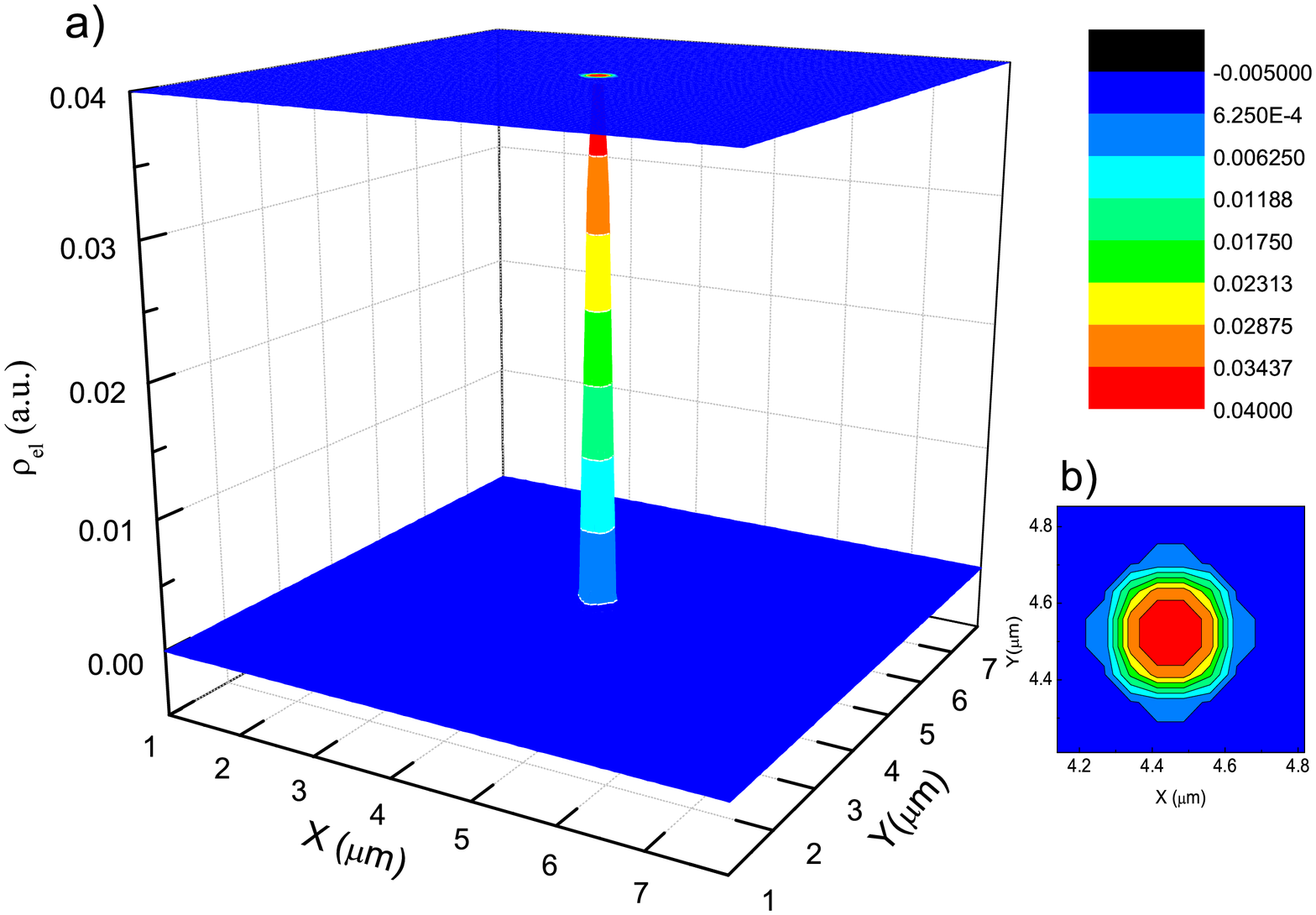}
\caption{(a) Charge distribution of 2DES with square gate model for the quantum
slab, the particle number $N$ in the quantum dot is 20, Sample-II.
(b) The top view of the charge distribution.} \label{fig:Q2DES-slab}
\end{figure}

\begin{figure}[h!] \centering
\includegraphics[width=0.7\columnwidth]{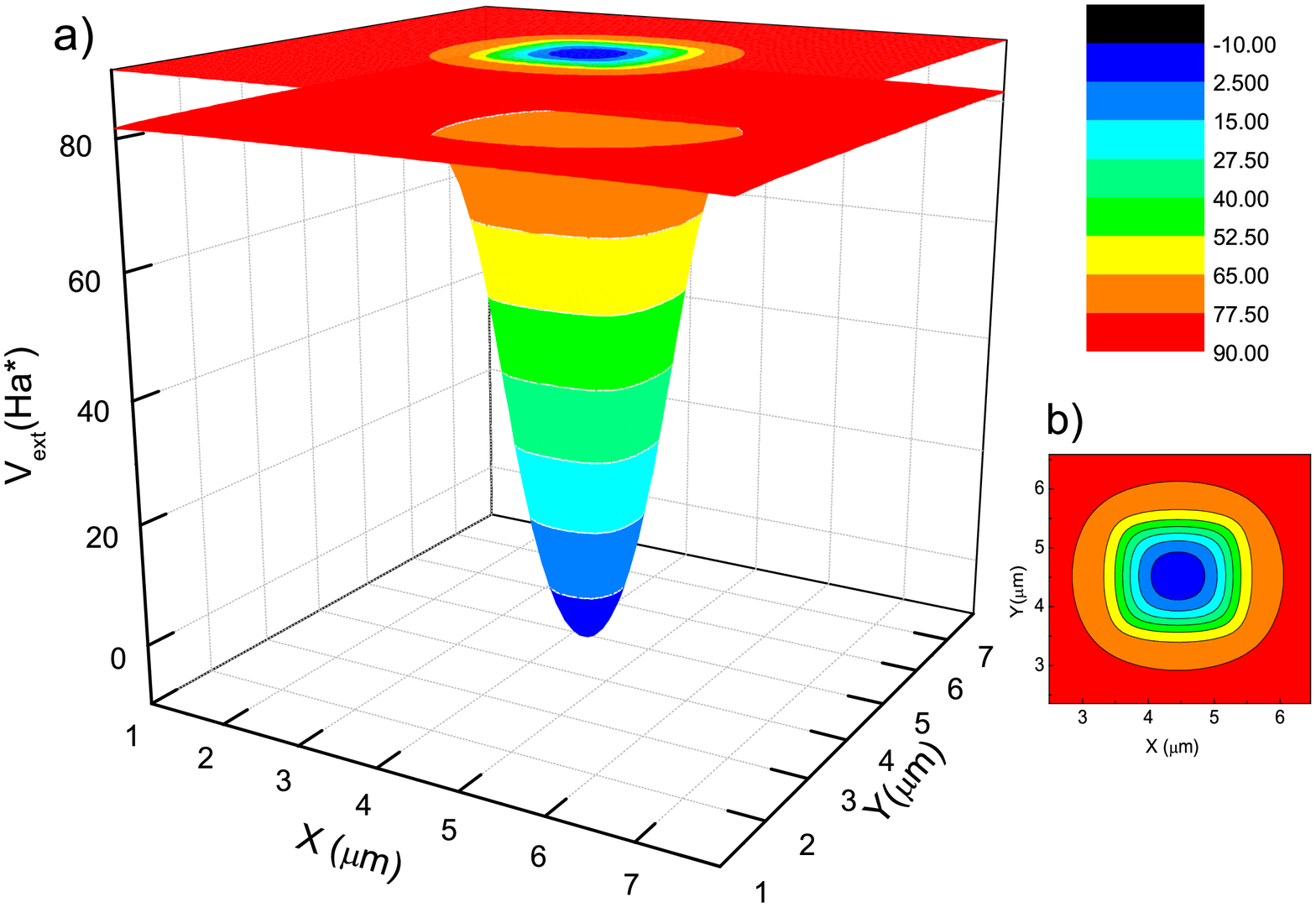}
\caption{(a) Potential profile of 2DES with square gate model
for the quantum slab, $N=20$, Sample-II (b) top view of potential profile.}
\label{fig:Vext-slab}
\end{figure}

\begin{figure} \centering
\includegraphics[width=0.9\columnwidth]{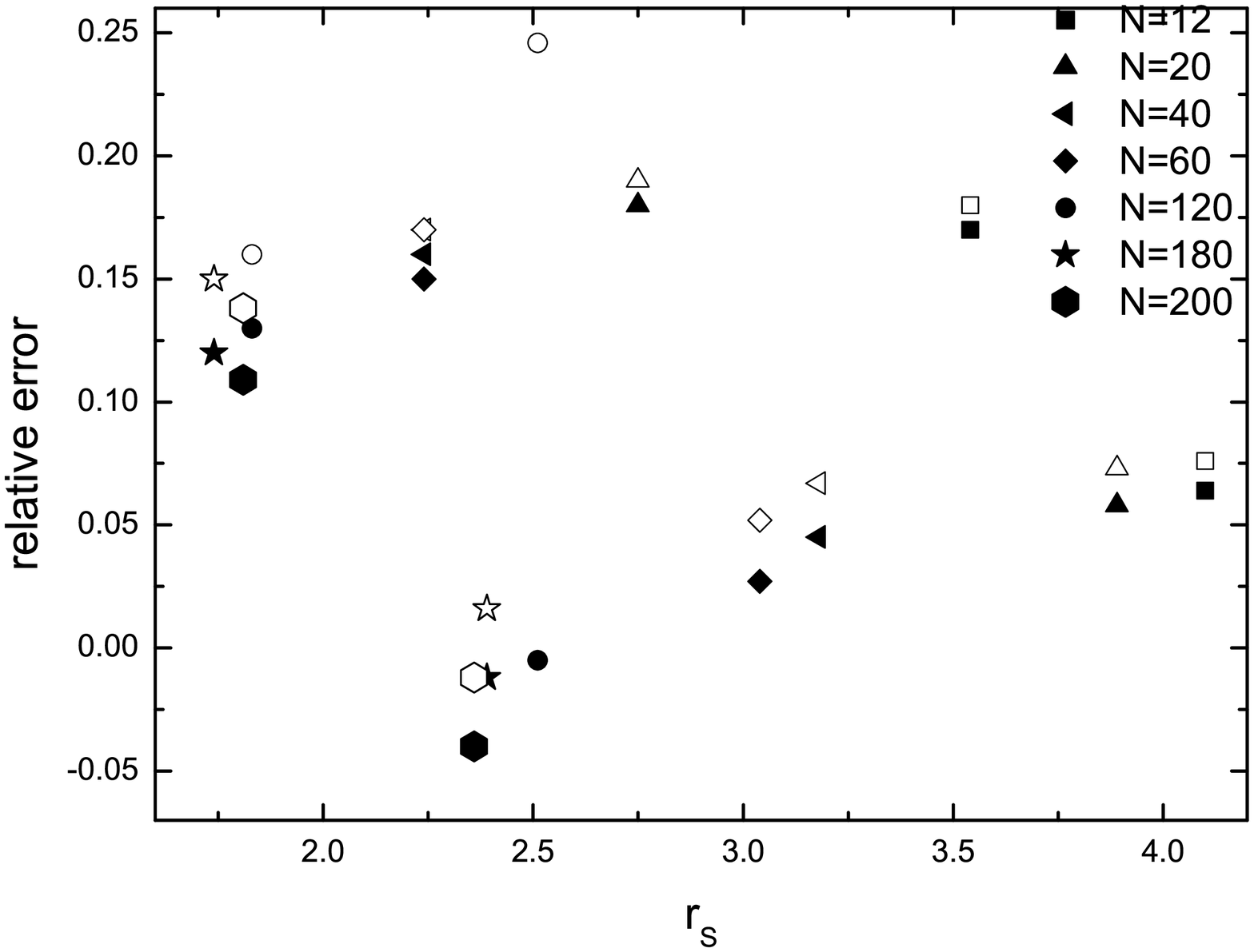}
\caption{Relative error ($|E_{LDA}-E|/E_{LDA}$) in the total energy
calculated within OFEF (filled symbols) and within the Thomas-Fermi
approximation (open symbols) for rectangular quantum slabs with
$N=12,...200$ as a function of the density parameter $r_s$.}
\label{fig:Slab-bagilHata-grafik}
\end{figure}

In our numerical simulations we consider a unit cell containing the
quantum dot which has the physical dimensions, $L_x=L_y=7.9\mu m$ by a
matrix of $128\times 128$ mesh points. As an example, in Figs.~\ref{fig:Q2DES-slab} and \ref{fig:Vext-slab}, we show the charge
distribution and the external potential profile only for $N=20$
particles in the QS.
Total energies of the QSs are calculated by using the potential and
charge distributions obtained via solving the Poisson equation,
self-consistently. The density parameter $r_s$  can be used to
define the average electron density in the quantum dot and this
parameter is determined from $r_s=\sqrt{A/\pi N}$ considering hard
wall boundary conditions for the quantum slabs, where $A$ is the
2DES's area. In the standard TF approximation the total energy is
given by
\begin{equation}\label{TF-enerji}
E[\rho]=T_{TF}[\rho]+\frac{1}{2}\int d\vec{r}\int
d\vec{r}'\frac{\rho(\vec{r})\rho(\vec{r}')}{|\vec{r}-\vec{r}'|}+\int
d\vec{r}\rho(\vec{r})v_{ext}(\vec{r}).
\end{equation}
Since the electron-electron interactions are taken into account only
classically, there is an important deficiency in Thomas-Fermi energy
functional. Therefore in limited particle and low density regime,
performance of this method is questionable due to the lack of the
quantum mechanical effects such as exchange and correlation
\cite{Pittalis09:165112}. In order to improve TF approximation, a
nonempirical, orbital-free energy functional for the total energy of
interacting electrons is proposed \cite{Pittalis09:165112}. This
functional is defined for the two-dimensional system as
\begin{equation}\label{OF-enerji}
E[\rho]=T_{TF}[\rho]+\frac{\pi}{2}\sqrt{\frac{N-1}{2}}\int
 d\vec{r}\rho^{3/2}(\vec{r})+\int
d\vec{r}\rho(\vec{r})v_{ext}(\vec{r}).
\end{equation}
It allows us to compute the total e-e interaction in a very simple
form. We calculate and compare the total energies of QSs within
orbital-free energy functional, Thomas-Fermi method
and LDA.\\
In Fig. \ref{fig:Slab-bagilHata-grafik}, we show the relative error
in the total energy calculated within OFEF and LDA for QS
considering various particle numbers. In Table I and II, the
energies obtained from TF, OFEF and LDA are shown. By using the
density and the potential profiles obtained from self-consistent
procedure, we calculate the total energy of electrons in QS. In this
work, we use the LDA energies obtained from \texttt{OCTOPUS}
real-space DFT code \cite{A.Castro2006} as the reference data.
External confinement potential is obtained by a numerical
interpolation utilizing a forth order polynomial fitting to the
self-consistent potential.
\begin{table}[h!]\center{
\begin{tabular}{|c|c|c|c|c|c|}
\hline
  N & $r_s$&  $ - E_{OFEF}$  &  $-E_{TF}$ & $-E_{LDA}$ & $\Delta
($\%$)$\\
  \hline
  12& 4.10&  56.12 & 55.41 & 59.99&  1.3\\
  \hline
  20 & 3.89& 102.92 & 101.28 & 109.25 & 1.6\\
   \hline
  40 & 3.18 & 237.08& 231.62 & 248.38 &  2.3 \\
\hline
  60& 3.04 & 401.70 & 391.35 & 412.92 & 2.6 \\
  \hline
  120& 2.51 & 991.11 & 962.05 & 986.32&2.9  \\
  \hline
  150& 2.39 & 1369.80 & 1331.40 & 1353.70& 2.8\\
  \hline
  200& 2.36 & 2001.10 & 1940.50 & 1917.38&3.0 \\
  \hline
\end{tabular}
\caption{Comparison of the total energy $E$ values (in effective
Hartree units) calculated within OFEF, TF approximation and LDA for
Sample-I. The QS is defined by trench-gating. The last column
defines relative errors,
 $\Delta=|(E_{TF}-E_{OFEF})/E_{OFEF}|$, considering $N$ particles.}
 \label{tab:a}}
\end{table}
%
%\vspace{2cm}
%
\begin{table}[h!]\center{
\begin{tabular}{|c|c|c|c|c|c|}
\hline
  N & $r_s$& $- E_{OFEF}$  &  $- E_{TF}$ & $- E_{LDA}$ &$\Delta
($\%$)$ \\
  \hline
  12& 3.54& 87.05 & 86.25 & 104.93&0.9\\
  \hline
  20 & 2.75& 153.83 & 151.86 &187.09 &1.3 \\
   \hline
  40 & 2.24 &363.15& 356.59 & 432.51 &1.8   \\
\hline
  60& 2.24 &605.81 &592.04 &713.45 &2.3  \\
  \hline
  120& 1.83 &1488.1 &1444.8 & 1715.66 &2.9  \\
  \hline
  150& 1.74 & 2031.5 & 1973.20 & 2322.85 & 2.9\\
  \hline
  200& 1.81 &3012.90 &2917.80 &3383.74&3.2 \\
  \hline
\end{tabular}
\caption{Comparison of the total energy $E$ (in effective Hartree
units) calculated within OFEF, TF approximation and LDA for
Sample-II, similar to Table.\ref{tab:a}} \label{tab:b}}
\end{table}
From Table I, it is seen that the energy differences are not
remarkably high for the QS up to $N=60$ particles, while the
differences are increasing for more number of particles. Meanwhile,
our calculations show that the corresponding energy differences
considering $N>60$ particles are also affected by the density
parameter $r_s$. We find a good agreement in the total energies
between LDA and OFEF. Despite the TF results are also close to the
reference data, the relative errors of orbital-free functional
remains below the TF.

\section{Conclusion}
In summary, we obtained the charge densities of QD systems by
solving the 3D Poisson equation. Next, we utilized the obtained
density considering a trench-gated structure to calculate the energy
of the system by using the orbital-free functional, TF and LDA.
According to our calculations, this new functional yields remarkably
accurate results in many electron systems in the limit of low
density and is very efficient due to its orbital-free form. Our
future aim is to employ this calculation scheme to quantum point
contacts and to investigate their transport properties. \vspace{1cm}
\ack \label{} We would like to thank Dr. Esa R\"{a}s\"{a}nen, for
introducing us the concepts of OFEF and for his critical reading of
the manuscript. This work was partially supported by the Scientific
and Technical Research Council of Turkey (T{\"U}B{\.I}TAK) under
grant no 109T083, IU-BAP:6970, DEU-BAP:2009183, and DEU-BAP:2009184.

\section*{References}
%\bibliography{zitate}

\begin{thebibliography}{10}
%
\bibitem{Jacak98:Book} L. Jacak, P. Hawrylak, and A.W$\acute{o}$js,
Quantum Dots (Springer, Berlin, 1998).
%
\bibitem{Parr_Gross-book-springer} R.G. Parr and W. Yang.
Density-functional Theory of Atoms and Molecules (Oxford University
Press, New York/Clarendon, Oxford,1989); R.M. Dreizler and E.K.U.
Gross, Density functional theory (Springer, Berlin, 1990).
\bibitem{Reimann02:1283} S. M. Reimann and M. Manninen, Rev. Mod.
Phys. 74, 1283 (2002).
\bibitem{Sakiroglu10:012505}
S. Sakiroglu and E. R\"{a}s\"{a}nen , Phys. Rev. A 82, 012505 (2010)
\bibitem{Rasanen10:195103}
E. R\"{a}s\"{a}nen, S. Pittalis, C. R. Proetto, Phys. Rev. B 81,
195103 (2010)
\bibitem{Rasanen10:115108}
S. Pittalis, E. R\"{a}s\"{a}nen and C.R. Proetto, Phys. Rev. B 81,
115108 (2010)
\bibitem{Gori-Giorgi09:166402}
P. Gori-Giorgi,M. Seidl and G. Vignale, Phys. Rev. Lett. 103, 166402
(2009)
\bibitem{Constantin08:016406}
L. A. Constantin, J. P. Perdew and J. M. Pitarke, Phys. Rev. Lett.
101, 016406 (2008)
\bibitem{Pittalis09:165112}  S. Pittalis and E. R\"{a}s\"{a}nen, Phys.Rev. B 80, 165112 (2009)
\bibitem{Wang00:117} Y. A. Wang and E. A. Carter, in Theoretical Methods in Condensed
Phase Chemistry, Progress in Theoretical Chemistry and Physics
Series, edited by S.D. Schwartz (Kluwer, Dordrecht,2000),
pp.117-184.
\bibitem{Ligneres05:springer} Ligneres V L and Carter E A, An Introduction to Orbital-Free
Density Functional Theory (Springer, Netherlands, 2005)
\bibitem{Watson00:128} S. C. Watson and E. A. Carter, Comput. Phys. Commun 128, 67 (2000)

\bibitem{Lieb95:10646} E.H. Lieb, J.P. Solovej, and J. Yngvason, Phys.
Rev. B 51, 10646 (1995).

\bibitem{Siddiki03:125315} A. Siddiki and R. R. Gerhadts, Phys. Rev. B 68 125315 (2003)

\bibitem{Siddiki07:045325} A. Siddiki and F. Marquardt, Phys. Rev. B 75, 045325 (2007).
\bibitem{Aslan-Tez} S. Arslan, M.S. thesis, Technical University of
Munich, 2008.
\bibitem{Aslan08:125423} S. Arslan, E. Cicek, D. Eksi, S.Aktas,
A.Weichselbaum and A.Siddiki, Phys. Rev. B 78, 125423 (2008).


\bibitem{Davies94:4800}J. H. Davies and I. A. Larkin, Phys. Rev. B 49, 4800 (1994).

\bibitem{A.Castro2006} A. Castro, H. Appel, M. Oliveira, C. A. Rozzi, X. Andrade, F.
Lorenzen, M. A. L. Marques, E. K. U. Gross, and A. Rubio, Phys.
Status Solidi 243, 2465 (2006) b.


\end{thebibliography}
\bibliographystyle{unsrt}
%\begin{thebibliography}{10}
%\bibitem{book1} Goosens M, Rahtz S and Mittelbach F 1997 {\it The \LaTeX\ Graphics Companion\/}
%(Reading, MA: Addison-Wesley)
%\bibitem{eps} Reckdahl K 1997 {\it Using Imported Graphics in \LaTeX\ } (search CTAN for the file `epslatex ')
%\end{thebibliography}

\end{document}